\documentclass[twocolumn,prl,showpacs,preprintnumbers,nofootinbib]{revtex4}
\usepackage{amsfonts}
\usepackage{graphicx}% Include figure files
\usepackage{graphicx}% Include figure files
\usepackage{dcolumn}% Align table columns on decimal point
\usepackage{amssymb}
\usepackage{amsmath}
\usepackage{epsfig}

\usepackage[colorlinks=true,linkcolor=red,citecolor=blue,urlcolor=cyan,
bookmarks=true,bookmarksopen=true,pdfpagemode=None,pdfstartview=FitH]{hyperref}
\newcommand{\be}{\begin{equation}}
\newcommand{\ee}{\end{equation}}
\newcommand{\bea}{\begin{eqnarray}}
\newcommand{\eea}{\end{eqnarray}}

\def\lapp{\mathrel{\rlap{\raise.5ex\hbox{$<$}}
                    {\lower.5ex\hbox{$\sim$}}}}
\def\gapp{\mathrel{\rlap{\raise.5ex\hbox{$>$}}
                    {\lower.5ex\hbox{$\sim$}}}}

\def\M0{m_{0}}
\def\MHF{m_{1/2}}
\def\MSL{m_{\tilde e_L}}
\def\MSTAU1{m_{\tilde \tau_1}}
\def\MXI10{m_{\tilde \chi_1^0}}
\def\MST1{m_{\tilde t_1}}
\def\MGL{m_{\tilde g}}
\def\TANB{\tan\beta}
\def\MHU{m_{H_u}^2}
\def\MHD{m_{H_d}^2}
\def\msum{\sum_i m_{\nu_i}}

\def\I{\mathrm{i}}

\begin{document}
\title {
The quark-lepton unification : 
LHC data and neutrino masses}
%\vskip .1in

\author
{\sf Abhijit Samanta\footnote{E-mail address: abhijit.samanta@gmail.com}}
\affiliation
{
{\em Department of Physics, Heritage Institute of Technology, Kolkata 700 107, India 
%\footnote{previous address: 
%Ramakrishna Mission Vivekananda University, Belur Math, Howrah 711 202,
%India}
}}
\date{\today}
%%%%%%%%%%%%%%%%%%%%%%%%%%%%%%%%%%%%%%%%%%%%%%%%%%%
\begin{abstract}
The recent discovery of nonzero $\theta_{13}$ (equal to Cabbibo angle $\theta_C$ 
up to a factor of $\sqrt{2}$),  
the masses of supersymmetric particles $\gapp$ TeV from LHC data,  and
the sum of three active neutrino masses $\sum_i m_{\nu_i}\lapp 1$ eV from
the study of large scale structure of the universe motivate to study whether
quark and lepton mixing have the same origin at the grand unification scale.
We find that both results from  neutrino experiments and LHC are complementary
in quark-lepton unified model. % (which indicates an unique structure of Yukawa matrix, close to unit matrix). 
A new constraint on SUSY parameters appears 
from electroweak symmetry breaking with a new correlation between the
lower bounds on sparticle masses and the upper bound on $\sum_i m_{\nu_i}$.
In addition, we find that only $\mu>0$ (which is favored by $(g-2)$ of muon) is allowed 
and $m_{\tilde q, \tilde \l} \gapp$ TeV if $\sum_i m_{\nu_i} \lapp 1$ eV.
On the other hand, a small change in  lower limit on $\theta_{13}$ 
from zero leads to a large increase in lower limits on sparticles masses ($\gapp 2$ TeV), 
which are  also the bounds if recently discovered boson  at LHC with mass around 125 GeV is the Higgs boson.   

\end{abstract}
%%%%%%%%%%%%%%%%%%%%%%%%%%%%%%%%%%%%%%%%%%%%%%%%%%%


\maketitle

\noindent
%%%%%%%%%%%%%%%%%%%%%%%%%%%%%%%%%%%%%%%%%%%%%%%%%
\section{Introduction} %\label{Introduction}
%%%%%%%%%%%%%%%%%%%%%%%%%%%%%%%%%%%%%%%%%%%%%%%%%
The standard model (SM) \cite{Pati:1974yy} of elementary particles is now 
a completely successful unified theory  with an answer to the origin of masses of quarks, 
leptons and  gauge bosons if the recently discovered new boson at Large Hadron Collider (LHC) 
at a mass around 125 GeV \cite{cmshiggs} is the standard model Higgs boson.
The present data on neutrino masses indicate see-saw scale equal or very close to the grand unification scale. 
%, the assumption (see-saw scale at the GUT scale) is a realistic one.  
Again, recent result after Daya Bay and RENO experiments \cite{dayabayreno} $\theta_{13} \approx \theta_C$ (equal up to a factor of $\sqrt 2$) possibly be a hint of a connection between leptonic mixing and quark mixing. The connection, particularly, the quark-lepton unification  has been first enunciated in the grand unified theories (GUTs) with an additional family symmetry  in \cite{rnm} and then worked out in a series of papers \cite{sfking}.  The family symmetry (e.g., $S_4$, $A_4$ and $SO(3)$ etc.) dictates the organizing principle for the structure of Yukawa matrix (to generate the observed mass pattern in both lepton and quark sector at weak scale).
%This motivates to study the grand unified theories (GUTs) 
%to find whether the origin of quark and the lepton mixings are the same.
%The unification between quarks and lepton is one of the key ingredient of the GUTs. 

The quark masses  originate from electroweak symmetry breaking (EWSB), while neutrino 
masses have different origin around the GUT scale --- the see-saw mechanism. 
All observed quark mixing angles are very small, while in neutrino sector 1-2 and 2-3
mixing angles are large and 1-3 mixing angle is small.  The  large magnifications of solar and atmospheric 
mixing angles through renormalization group evolution (RGE) from  a high scale around the
GUT scale to weak scale are possible \cite{Antusch:2003kp} for quasi-degenerate 
neutrinos. Here, % Yukawa matrix close to the unit matrix  is needed to generate 
the required quasi-degenerate 
%%%%%%%%%%%%%%%%%%% PLB revision %%%%%%%%%%%%%%%%%%%%%%%%%%%%%%%%%
hierarchical (normal) 
%%%%%%%%%%%%%%%%%%%%%%%%%%%%%%%%%%%%%%%%%%%%%%%%%%%%%%%%%%%%%
neutrinos can be generated in GUT models with type II see-saw  with an additional family symmetry \cite{sfking} (discussed later).

The intermediate scale between GUT scale and electroweak (EW) scale is relevant for see-saw mechanisms, 
but not necessary at all for quark masses and their mixing. 
The lower limits of the see-saw
scales are increased  as the neutrino masses becomes smaller and smaller. The improvement on
sky survey data are showing  more stronger lower limit on the sum of neutrino masses $\sim$ 0.6 eV or more smaller,
which leads to see-saw scales equal or very close to the GUT scale.
If the see-saw scale is at the GUT scale, the RG evolved neutrino masses 
at the EW scale are fitted well over the allowed ranges obtained from sky survey data
\cite{dePutter:2012sh} as well as
the mass squared differences from neutrino oscillation experiments. 
%
%
%They can be obtained at high scale theories
%using dimension-5 operator \cite{Chankowski:2000fp}. 
%The present neutrino data require  the see-saw scale around the GUT scale.
%The energy dependence of the effective neutrino mass matrix 
%below the scale where this operator is generated, is described by its
%renormalization group equation (RGE) \cite{Antusch:2003kp}.
%
%
%The required values of the neutrino masses can be tested at experiments,
%as the weak scale values of these masses to explain the  
%oscillation data are in the range obtained from  sky survey data
%\cite{Hamann:2010bk} as well as from beta decays. 
Moreover, this range can also be accessible in future double beta decay 
experiments \cite{Barabash:2006fw} and in KATRIN experiment \cite{Osipowicz:2001sq}.

%The above experiments on neutrino masses in conjunction with LHC can verify  
%the grand unified theories where  the mixing matrices for quarks and leptons 
%are same at the grand unification scale (where the gauge couplings unify).

%The unification of fundamental interactions at the GUT scale requires SM embedded
%in higher theories.  The supersymmetry is one of the most appealing candidate 
%beyond SM.

%Here, we point out an unique  structure of Yukawa matrix (close to unit matrix) at the GUT scale (required by quark-lepton unification) among many %
%structures dictated by the different symmetry groups, which not only  
%satisfies and/or predicts all experimental results available till now, but also shows that {\it both results from  neutrino experiments and LHC are 
%complementary}.
%In fact, after the discovery of the grand unified theory (tested by different  experiments), it will predict an unique structure of Yukawa matrix 
%at GUT scale, which should emerge from some underlying symmetry. The recent advancement of experimental results provides the scope to study the %%
%possible structure of the Yukawa matrices and it may be confirmed in future.

%%%%%%%%%%%%%%%%%%%%%%%%%%%%%%%%%%%%%%% PLB revision %%%%%%%%%%%%%%%%%%%%%%%%%%%%%%%%%%%%%%%%%%%%%%
Here, we consider the hypothesis of equal quark and lepton mixings and hierarchical (normal) neutrino masses at GUT scale as in \cite{rnm}.  In this scenario we find a correlation between the upper limit on the sum of the active neutrino masses ($\msum$) and lower bound on $\tan\beta$  as the quark-lepton unification fixes the lower limit on $\tan\beta$ (discussed later) for a given upper limit on $\msum$.  
%%%%%%%%%%%%%%%%%%%%%%%%%%%%%%%%%%% PLB revision ends %%%%%%%%%%%%%%%%%%%%%%%%%%%%%%%%%%%%%%%%%%%%%%%%%%%%
Once lower limit on $\tan\beta$ is fixed, lower bounds on sparticle masses are fixed from EWSB condition. It determines the stable the electroweak symmetry breaking minima  and fixes $\mu^2$:
 \be\mu^2 =\frac{m_{H_d}^2 - m_{H_u}^2\tan\beta}{\tan^2\beta -1} -\frac{1}{2}M_Z^2. \ee
If $\mu^2 < 0$, Higgsino mass is imaginary and  the EWSB minima becomes unstable.
The RGE of Higgs mass parameters $m^2_{H_u}$ and $m^2_{H_u}$ strongly depends on the sparticle masses and $\tan\beta$. 
This leads to
a correlation between lower limit on $\tan\beta$ and lower bounds on  
sparticle masses at weak scale. 

For $\sum_i m_{\nu_i} \le 1$ eV (constrained from large scale structure of universe), sparticle masses are $\gapp 1$ TeV
(consistent with LHC bounds) and only $\mu > 0$ is allowed (supported by $(g-2)$ of muon). Again, a small change in  lower limit on $\theta_{13}$ from zero ($\approx \theta_C/\sqrt{2}$ after Daya Bay and RENO experiments) leads to a large increase in lower limits 
on sparticles masses ($ \gapp 2$ TeV) from quark-lepton unification, which is  
also the case if recently discovered boson  at LHC with mass around 125 GeV is the Higgs boson.   
We present our result for constrained
minimal supersymmetric extension of the SM (CMSSM) 
\cite{cmssm}. %(commonly used by experimental groups ATLAS, CMS etc.) 
%
%
%Here, we consider the simplest supersymmetry (SUSY) breaking  model. 
%The important and new feature of this work is that 
In our study the running 
of  neutrino parameters are exact as they are coupled with 
the running of minimal supersymmetric standard model (MSSM) parameters 
using the ISASUGRA program of 
the ISAJET package (V7.81) \cite{Paige:2003mg}.
%The  supersymmetric contributions were approximate 
%in the previous works discussed earlier. 
%Here, we  consider two loop RGE for Yukawa, gauge and 
%sparticle masses, and one loop RGE for neutrino parameters.
%
%We achieve a new constraint on SUSY 
%parameters in quark-lepton unified models from 
%exact running of  neutrino masses and  mixing angles coupled with 
%the running of supersymmetric parameters.
%Here, larger values of  $\tan\beta$ is needed for quark-lepton unification, which 
%constrain SUSY parameter space from EWSB. 
  
\section{Models for degenerate neutrino masses at high scale}
%%%%%%%%%%%%%%%%%%%%%%% PLB %%%%%%%%%%%%%%%%
The most natural way to understand the smallness of neutrino mass is the see-saw mechanism. Here, the neutrino mass matrix is generated by the effective dimension 5 Weinberg operator \cite{Weinberg:1979sa}.
In conventional type I see-saw \cite{see-saw-I},  the SM is extended with additional heavy right handed neutrinos 
(which are not connected with the left handed fermions) and the mass matrix  can be  written as %\be M_\nu = - M_D (f v_R)^{-1} M_D^T\ee
\be M_\nu %= - \frac{v^2}{2} Y_N^T M_N^{-1} Y_N 
= - M_D (M_N)^{-1} M_D^T; \ee
where, % $f$ is the Yukawa coupling to the right handed neutrinos, $v_R$ is the $B-L$ symmetry breaking scale, 
$M_N$ is the mass matrix of right handed neutrinos
and $M_D$ is the Dirac neutrino mass matrix. Here, the neutrino masses are expected to be hierarchical in the   similar way to the quark masses. 

However, in case of type II see-saw \cite{see-saw-II} the SM model is extended by % right handed neutrinos %$\nu_R \sim (1,0)_{SU(2)\times U(1)}$and a 
a charged Higgs triplet $\Delta$ and %\sim (3,1)_{SU(2)\times U(1)}$, 
the neutrino mass matrix can be expressed as   
% %%%%%%%%%%%%%%%%%%%%%%%%%%%%%%%%%%%%%%%%%%%%%%%%%%%%% PLB revision ends %%%%%%%%%%%%%%%%%%%%%%%%%%%%%%%%%%
\be M_\nu = %\frac{v^2}{2} \frac{\Lambda Y_\Delta}{M_\Delta^2} =
 Y_\Delta <\Delta>, \ee
%where, $\Lambda$ is the coupling of the Higgs triplet with 
%
Now, the most general neutrino mass matrix can be written as 
\be M_\nu = Y_\Delta <\Delta > -  M_D (M_N)^{-1} M_D^T \label{e:see-sawII}\ee
The Yukawa coupling matrix $Y_\Delta$ depends on high scale physics and  it is unconstrained by SM data. One can therefore choose it to be unit matrix. If one considers the neutrino mass matrix dominated by the first term, then the neutrino masses are quasi-degenerate 
 in conjunction with the lepton mixing angles close to quark mixing angles. 
%It sh RGE in eqs. \ref{e:t} derived for dimension 5 Weinberg operator are valid as long as vacuum expectation value of the triplet Higgs responsible for the first term of eq. \ref{e:see-sawII} is heavier than the 

The realization of above type II see-saw scenario % (to generate quasi-degenerate neutrino masses and quak-leptopn unification) 
has been shown to be achieved in GUT models with gauge group $SU(2)_L\times SU(2)_R\times SU(4)_C$   and with an additional $S_4$ global symmetry in \cite{rnm}. This is not an adhoc assumption, while the gauge group can be a subgroup of number of GUTs like $SO(10)$, $E_6$, $SO(18)$ etc. 

Here, we have not considered the running of the parameters between two scales  for decoupling of heavy fields as the values of neutrino parameters are considered as the input at the lowest scale of decoupling of Heavy fields and  this lowest scale
is assumed as the GUT scale.  

\noindent
%%%%%%%%%%%%%%%%%%%%%%%%%%%%%%%%%%%%%%%%%%%%%%%%%
\section{Renormalization group evolution}
%%%%%%%%%%%%%%%%%%%%%%%%%%%%%%%%%%%%%%%%%%%%%%%%%
The solution of the coupled RGEs for neutrino parameters 
along with SUSY parameters are obtained by an iterative cyclic  process 
(weak-to-GUT and then GUT-to-weak)  with GUT
boundary conditions  following CMSSM \cite{cmssm}.   
The neutrino parameters are also set at GUT scale.
The Higgsino mass $\mu$ and the soft  Higgs bilinear term $B$ are fixed from radiative 
electroweak symmetry breaking (REWSB). 
This has been done by the following steps.
First, we set  the gauge couplings and Yukawa couplings
at EW scale and run only these couplings  up to GUT scale
(where $g_1$ and $g_2$ meet) setting  other required mass parameters 
at SUSY breaking scale with approximate values. 
Now, we set the GUT boundary conditions for neutrino 
and SUSY parameters; put $\mu, ~ B = 0$ (one can also put arbitrary values), 
and run down to weak scale.
After adding the loop corrections to  
$\MHU$ and $\MHD$ 
we calculate  $\mu$ and  $B$ from EWSB condition.
Taking these $\mu$ and $B$ values as well as the RGE evolved
SUSY parameters and neutrino parameters, we run up to 
GUT scale and put the GUT values of $\mu$ and $B$ as they come
and reset all other parameters at the GUT scale as earlier.
We iterate this process  until all parameters converge to a certain 
tolerance. The iteration are needed as the $\mu$ and $B$ are
involved in the running of other parameters.

%%%%%%%%%%%%%%%%%%%%%%%%%%%%%%%%%%%%%%%%%%%%%%%%%
%\subsection{Renormalization group equations for neutrino masses and mixing angles}
%%%%%%%%%%%%%%%%%%%%%%%%%%%%%%%%%%%%%%%%%%%%%%%%%
To understand the results the analytical formula for RGE 
of neutrino mixing angles and masses \cite{Antusch:2003kp}
 are very important:
\noindent
\begin{subequations}\label{e:t}
\bea
\Dot{\theta}_{12}
& = & 
        -\frac{C y_\tau^2}{32\pi^2} \,
        \sin 2\theta_{12} \, s_{23}^2\, 
        \frac{
      | {m_1}\, e^{\I \varphi_1} + {m_2}\, e^{\I  \varphi_2}|^2
     }{\Delta m^2_\mathrm{21} }     
\nonumber\\
 & &       + {O}(\theta_{13}) \;,
         \label{e:theta12dot}
\eea
\bea
\Dot{\theta}_{13}
 &=&  
        \frac{C y_\tau^2}{32\pi^2} \, 
        \sin 2\theta_{12} \, \sin 2\theta_{23} \,
        \frac{m_3}{\Delta m^2_\mathrm{32} \left( 1+\zeta \right)}
        \times
\nonumber\\
&& 
\times
        \left[
         m_1 \cos(\varphi_1-\delta) -
         \left( 1+\zeta \right) m_2 \, \cos(\varphi_2-\delta) -
\right .
\nonumber\\
&&
\left .
        \zeta m_3 \, \cos\delta
        \right ]
        +       {O}(\theta_{13}) \;,
 \label{e:theta13dot}
\eea
\bea 
        \Dot{\theta}_{23}
 &=& 
        -\frac{C y_\tau^2}{32\pi^2} \, \sin 2\theta_{23} \,
        \frac{1}{\Delta m^2_\mathrm{32}} 
%\nonumber\\
%&&
        \left[
         c_{12}^2 \, |m_2\, e^{\I \varphi_2} + m_3|^2 
\right .
\nonumber \\
&&
\left .
       + s_{12}^2 \, \frac{|m_1\, e^{\I \varphi_1} + m_3|^2}{1+\zeta}
        \right] 
%        \nonumber\\
%        & & {}
        + {O}(\theta_{13}) \;.
 \label{e:theta23dot}
% \nonumber\\
\eea
\end{subequations}
where
$\zeta=\Delta m^2_\mathrm{21}/\Delta m^2_{32}$, $\Delta m^2_{21} = m_2^2-m_1^2,~ 
\Delta m^2_\mathrm{32}= m_3^2-m_2^2$; 
$C=1$ in MSSM
and $-3/2$ in SM. 
%\begin{subequations}\label{e:m}
%\bea
% 16\pi^2\,\Dot{m}_{1}
% & = &
%       \left[
%        \alpha + C y_\tau^2 \left( 2 s_{12}^2 \, s_{23}^2 + F_1 \right)
%        \right] m_1 \;,
%\\
% 16\pi^2\,\Dot{m}_2
% & = &
%        \left[
%        \alpha + C y_\tau^2 \left( 2 c_{12}^2 \, s_{23}^2 + F_2 \right)
%        \right] m_2
% \;,
% \\
% 16\pi^2\,\Dot{m}_3
% & = &
% \left[ \alpha
% +2 C y_\tau^2 \, c_{13}^2 \, c_{23}^2
% \right] m_3
% \;,
%\eea
%\end{subequations}
%where $\alpha=-3g_2^2+2y_\tau^2+6(y_t^2+y_b^2) +\lambda$ for SM;
%and  $\alpha=-6/5g_1^2-6g_2^2 +6y_t^2$ for MSSM. We are using GUT
%charge normalization for $g_1$.
% \(F_1\) and \(F_2\) contain terms proportional to $\sin\theta_{13}$:
%\begin{subequations}\label{eq:Fi}
%\begin{eqnarray}
%        F_1 &=&
%        -s_{13} \, \sin 2\theta_{12} \, \sin 2\theta_{23} \, \cos\delta +
%        2 s_{13}^2 \, c_{12}^2 \, c_{23}^2 \;,
%\\
%        F_2 &=&
%    s_{13} \, \sin 2\theta_{12} \, \sin 2\theta_{23} \, \cos\delta +
%        2 s_{13}^2 \, s_{12}^2 \, c_{23}^2 \;.
%\end{eqnarray}
%\end{subequations}
%

From the RGE it is clear that  large value of $y_{\tau}$ (which requires large $\tan\beta$), quasi-degenerate neutrino masses, 
and normal hierarchical mass pattern are needed to generate large radiative magnification
of the 1-2 and 2-3 mixing angles at electroweak scale from quark-lepton unified mixing angles at the GUT scale.  
%
%The large magnifications of solar and atmospheric mixing 
%angles at weak scale for certain choices of the 
%parameters was realized in \cite{
%Mohapatra:2003tw, Mohapatra:2005pw,
%Balaji:2000au, Antusch:2002fr} %Antusch:2005gp
%through  
%renormalization group evolution (RGE) in supersymmetry (SUSY) 
%embedded theory. 
%But, very large threshold corrections (one order magnitude larger) 
%are needed to fit the experimentally measured values 
%of the mass squared differences in absence of Majorana phases, 
%which requires a special kind of model with inverted 
%slepton mass pattern: 
%$m_{\tilde e}/m_{\tilde \mu,\tilde \tau} = 1.3 -2.5$
%\cite{Chankowski:2000fp, Mohapatra:2005gs}.
%
%In presence of the Majorana phases 
%both  neutrino mass squared differences as well as  mixing angles 
%obtained from  oscillation data can  naturally be generated 
%through the renormalization group (RG) running 
%without requiring any threshold corrections \cite{Agarwalla:2006dj}.

%%%%%%%%%%%%%%%%%%%%%%%%%%%%%%%%%%%%%%%%%%%%
\section{Result}
%%%%%%%%%%%%%%%%%%%%%%%%%%%%%%%%%%%%%%%%%%%%%%%%%
The sparticle masses are complicated function of soft SUSY breaking parameters which are 
obtained at weak scale through running of their coupled RGE from GUT scale.   
We find the allowed parameter space (and lower limits on masses)  by  scanning 
randomly over the following ranges of the parameters; common scalar mass 
($\M0$): 0.05 - 3 TeV, common gaugino mass ($\MHF$): 0.05 - 3 TeV,
common trilinear coupling ($A_0$): $-3\M0 - +3\M0$, sign($\mu$): $\pm 1$,
and $\tan\beta$ (ratio of two vacuum expectation values of $H_u$ and $H_d$): 35 - 70. 
We set the ranges for neutrino masses $m_1^0,~m_2^0,~m_3^0=  0 - 0.7$ eV, 
neutrino mixing angles
$\theta_{12}^0=0.22(1\pm x_1),~ 
\theta_{13}^0=0.0039(1\pm x_2),~ 
\theta_{23}^0=0.034(1\pm x_3)$,
CP phase $\delta_{CP}^0=60^\circ(1\pm x_4)$,  and 
Majorana phases $\varphi_1^0, ~ \varphi_2^0 = 0-360^\circ$.
We have chosen $x_i$s (uncertainties in the unification) randomly within the range 
$0\le x_i\le 40\%$ ($i=1,2, ...$)  as an uncertainty due to approximate evolution
\cite{Naculich:1993ah}
of CKM parameters.
We have checked varying the upper limits of  $x_i$s from 30\% to 50\% that the results
do not change drastically; the bounds (discussed later)
become gradually stronger as the upper limits of $x_i$s are decreased.   

We set the experimental bounds obtained from LEP data: $m_h > 114.5$ GeV, 
$m_{\tilde\chi^\pm} > 103$ GeV \cite{LEP} as these masses can be
dominated by the value of $\mu$.
% which is very small at high 
%$\tan\beta$
If we withdraw the LEP bounds, the relatively lower 
sparticle masses are allowed,  but the correlation of 
the bounds (discussed later) with $\msum$ remains.
We consider only the points in the parameter space that
can produce 
neutrino oscillation parameters at weak scale within the 
3$\sigma$ range obtained from global fit \cite{Schwetz:2011zk}:
%$|\Delta m^{2}_{32}|\simeq 2.50^{+0.26}_{-0.36} %(^{+0.37}_{-0.17}) 
%\times 10^{-3}$ eV$^2$, 
%$\Delta m^{2}_{21} \simeq 7.59^{+0.60}_{-0.50}\times 10^{-5}$ eV$^2$,
               $\sin^2\theta_{23}\simeq$ ${0.52}^{+0.12}_{-0.13}$;
%while solar neutrinos tell us 
                 $\sin^2\theta_{12}\simeq {0.312}^{+0.048}_{-0.042}$, and 
                 $\sin^2\theta_{13}\lapp 0.039$ (the case for nonzero bound on $\theta_{13}$ has been discussed later). 
%{0.013}^{+0.022}_{-0.012}$. 
%%%%%%%%%%%%%%%%%%%%%%%%%%%%%%% PLB revision %%%%%%%%%%%%%%%%%%%%%%%%%%%%%%%%%%%%%%
It is expected that the addition of threshold corrections  to the mass squared differences 
\cite{Agarwalla:2006dj} may restrict the parameter space as it restricts  the choices of  $m_{\nu_i}$s for a given $\msum$. 
We have studied the unification with and without threshold corrections varying the ranges of mixing angles 
and mass squared differences.   But, these are not very significant in this scenario with normal slepton mass hierarchy and no 
significant change in result is observed as we have used large allowed  ranges for all parameters in our scanning (both neutrino parameters as well as SUSY parameters), where the parameters are chosen randomly. 
%%%%%%%%%%%%%%%%%%%%%%%%%%%%%%%%%%%%% PLB revision ends %%%%%%%%%%%%%%%%%%%%%%%%%%%%
We present the plots for 
$\Delta m^{2}_{21}$: $5-10\times 10^{-5}$ eV$^2$ and 
 $\Delta m^{2}_{32}$:
 $2-3\times 10^{-3}$ eV$^2$, respectively, 
considering the threshold corrections.
Obviously, more stronger constraints are  obtained for  narrower
ranges of oscillation parameters (discussed later).
 
%\footnote{

%We have checked that the results
%do not change significantly with the change in the range of mixing angles.  
%However, there is relatively strong dependence on  $\Delta m^{2}_{21}$
%than  $\Delta m^{2}_{32}$. 
%The change in the results can be easily understood in absence
%of threshold corrections. 
%If we increase $\Delta m^{2}_{21}$, the bounds begin 
%to be weaker, which is determined by the $\tau$ Yukawa coupling ($y_\tau$),
%and hence mainly by $\TANB$ (see detailed discussion later).    
%In case of absence of threshold corrections, we observed 
%that if $\Delta m^{2}_{21} > 9\times 10^{-5}$ eV$^2$
%the lower bounds of sparticle masses become significantly weaker. 
%The bounds are relatively more stable with threshold corrections with 
%the increase in $\Delta m^{2}_{21}$.  

%We should note here that then RGE is able to produce 
%whole ranges of the oscillation parameters, (particularly,
%we check $\theta_{13}$) from the above GUT 
%boundary condition. 

\begin{figure*}[htb]
\includegraphics[width=7.8cm,angle=0]{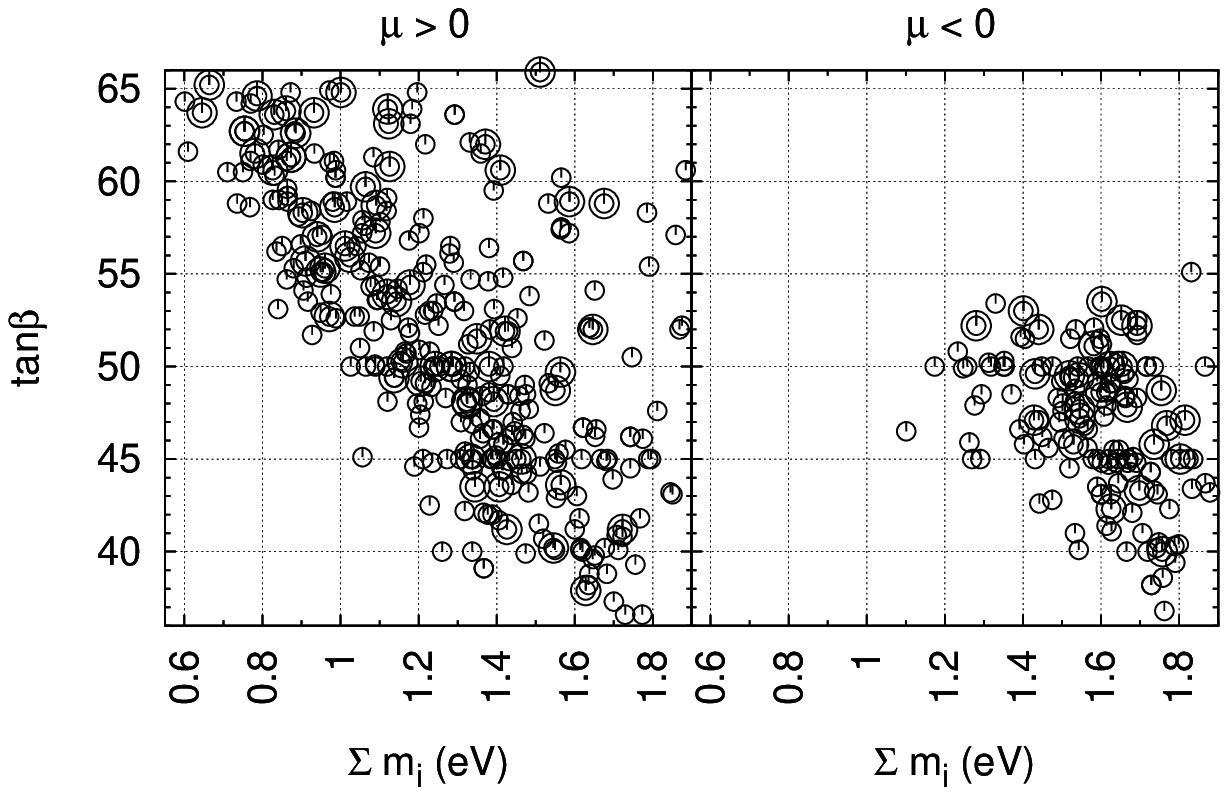}
\includegraphics[width=7.8cm,angle=0]{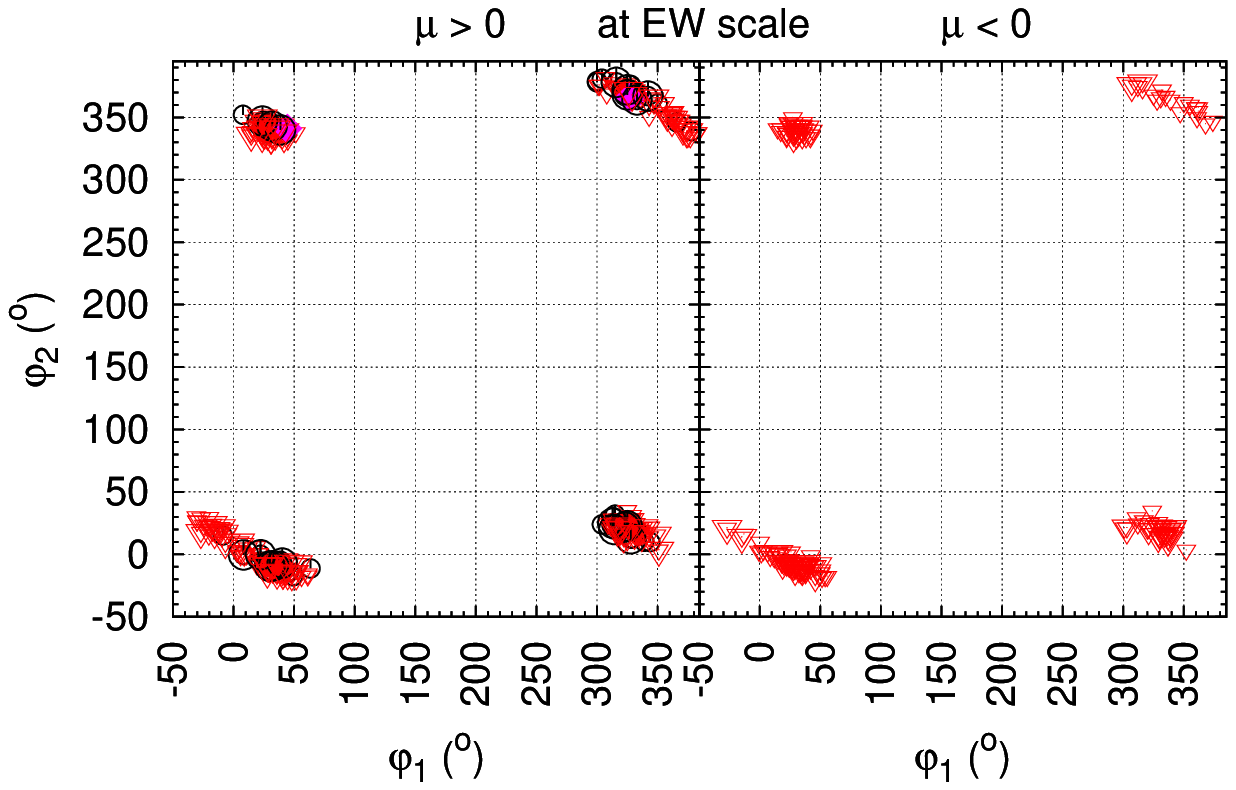}
\caption{\sf \small
The lower bound on $\tan\beta$ for a given $\msum$, 
and the allowed parameter space for Majorana phases in $\varphi_1-\varphi_2$ plane 
at the weak scale 
 for both sign of $\mu$.  
}
\label{f:tanb}
\end{figure*}

The generation of neutrino mixing angles at EW scale 
in the ranges allowed by global-fit of neutrino oscillation data  needs large 
radiative magnifications and demands very high value of $y_\tau$.  As the ranges  
of the mixing angles at present are very narrow, it almost fixes 
$y_\tau$  and consequently  determines the lower bound 
on $\tan\beta$ for a given ${\sum} m_{\nu_i}$ at the EW scale. 
As $\msum$ is lowered, higher value of $y_\tau$ is required and it demands
more larger value of $\tan\beta$. This is shown in the first plot of Fig. \ref{f:tanb}.  

The solar and atmospheric mass squared differences  are different by two order of magnitude 
as well as the magnification for solar angle is $\sim 3$ and for atmospheric angle is 
$\sim 20$. To accommodate all  parameters  in the experimentally allowed ranges  
for a given neutrino mass scale %$\msum=(m_1+m_2+m_3)/3$ 
the Majorana  phases are constrained in  very narrow  regions 
(see Fig. \ref{f:tanb}). 
This can be understood from  Eq. \ref{e:theta12dot} and from  Eq. \ref{e:theta23dot}.

The upper bound on $\tan\beta$ is either fixed from REWSB 
% (which determines the stable minima of Higgs potentialand fixes $\mu^2$) 
or from the LEP bounds on $m_{\tilde\chi^\pm}$ or $m_h$ (which are lowered
for smaller $\mu$ values). 
As $\TANB$  increases $\MHD$ decreases through RG evolution; and it 
can even be negative. This leads to smaller $\mu$ at larger $\TANB$.
At more higher $\TANB$, $\mu^2$ becomes negative as the minima of Higgs potential become unstable and then REWSB becomes impossible.  
In case of  $\mu < 0$, the loop correction to $\MHD$ leads to a more lower value compared to $\mu > 0$ and 
we find an upper limit on $\TANB \lapp 55$ from REWSB.
This restricts the increase in $y_\tau$ and consequently
leads to a lower bound on $\msum \approx 1$ eV,
which is  strongly disfavored  by the present 
cosmological data \cite{dePutter:2012sh}.
On the other hand, for $\mu >0$ one can increase 
$\TANB$ up to 65 leading to a decrease in $\msum\approx 0.6$ eV, 
which is very highly favored by  sky survey data \cite{dePutter:2012sh}.

The REWSB and the value $\mu^2$ depends on the GUT scale values of $\MHU$ and $\MHD$ 
as well as on other soft mass parameters (which change the values of $\MHU$ and $\MHD$ through RGE and loop corrections).  
Again, for a given $\TANB$,  $y_\tau$ depends on the value of $\mu$, gaugino mass parameters and
scalar mass parameters through radiative corrections at EW scale.  
This leads to strong  lower bounds on sparticle masses correlated with the 
upper limit of $\msum$.
If $\mu > 0$ and 
$\msum \le 1$ eV, then we find
\noindent
$m_{\tilde g} \gapp$ 1 TeV,   $m_{\tilde t_1} \gapp$ 0.7 TeV,  
and  $m_{\tilde e_L} \gapp$ 0.5 TeV. 
%$m_{\tilde g} \gapp 1 $ TeV for $m_\nu^{\rm avg} \lapp 0.4$ eV,
% $\MSL \gapp$ 0.5 TeV, and $\MGL \gapp$ 1 TeV. 
%
The lower bounds on sparticle masses depend only on upper limit on $\msum$ as other parameters are scanned
over their whole ranges. 
%
%The negative result
%of the search for SUSY by ATLAS collaboration may be a hint for
%unification of quark-lepton mixing as  
All these bounds  follow
the  recent LHC results \cite{cmshiggs}; and again,  
the neutrino mass limit $\msum \lapp 1$ eV is strongly favored 
by  the sky survey data.

\begin{figure*}[htb]
\includegraphics[width=7.8cm,angle=0]{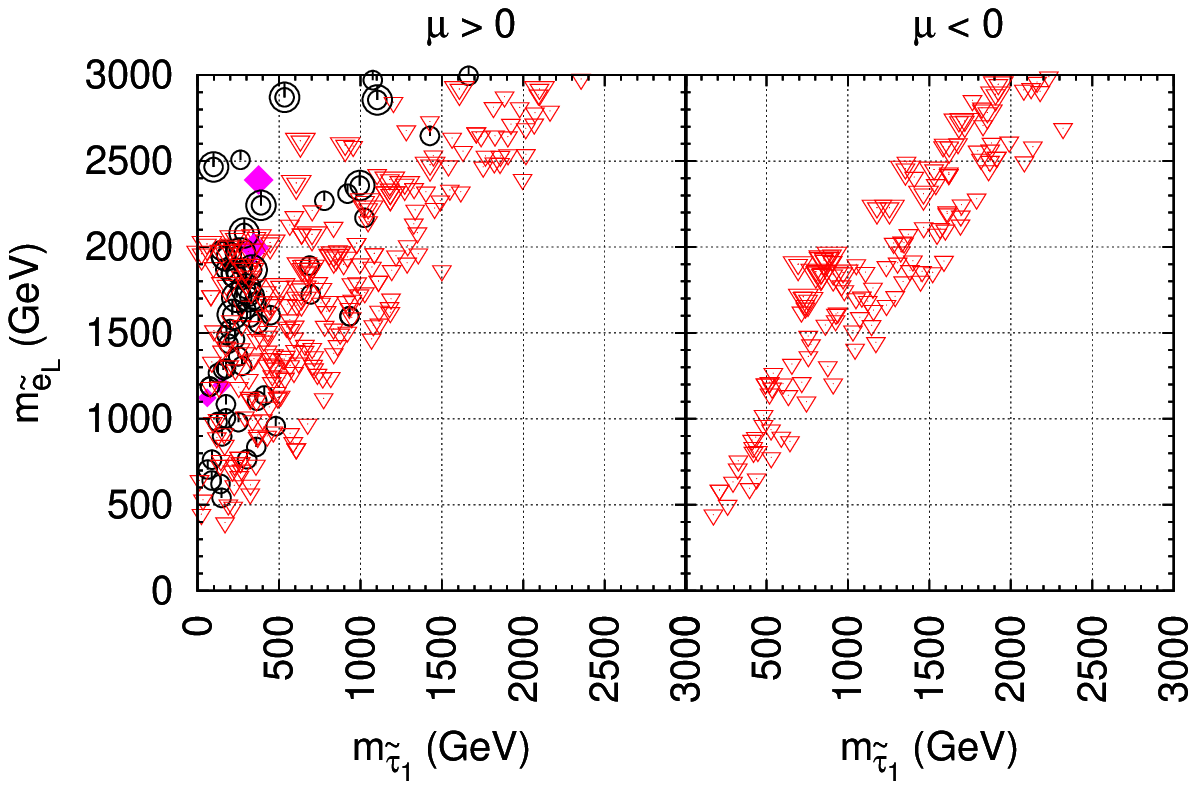}
\includegraphics[width=7.8cm,angle=0]{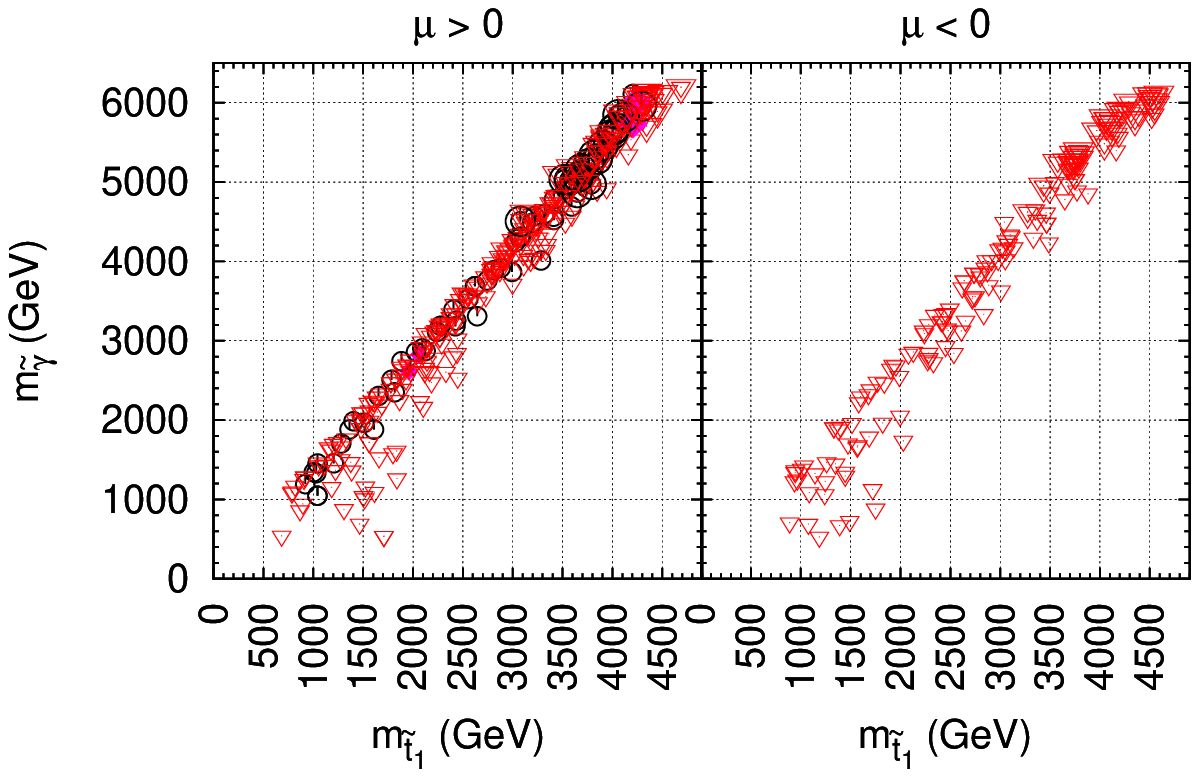}
\caption{\sf \small
The  allowed region in the plane of   %$\M0 - \MHF$ (first),
 $\MSL-\MSTAU1$,  %$\MSTAU1-\MXI10$ (third) 
and $\MGL-\MST1$, respectively.
The allowed points for $\msum <$ 0.7 eV,  0.7 - 1 eV
and $>$ 1 eV are represented by solid box, circle and triangle, respectively.
The bigger point-size denotes the points with further constraints on Higgs mass within 
the range 125 $\pm$ 1 GeV.  The shapes of the points signify the same as described earlier.  
}
\label{f:smass}
\end{figure*}

In  Fig. \ref{f:smass}  we present the allowed parameter space in
the planes of %$\M0-\MHF$, 
$\MSL-\MSTAU1$
%$\MSTAU1-\MXI10$, 
and $\MGL-\MST1$, respectively.
For each plot  the allowed points are separated  for three  
 $\msum$  ranges: $<$ 0.7 eV, 0.7 - 1.0 eV and $>$1 eV, respectively,  
to show the dependence of lower bounds of sparticle masses on the upper bound of $\msum$.
As an example,  in case of $\mu >0$,
we find $\M0 \gapp$ 1.5 TeV only when $\MHF \gapp 0.4$ TeV
for whole range of $\msum$ ($0.6 - 2$ eV); 
but, if $\msum \lapp 1$ eV, then   $\MHF \gapp 0.5$ TeV
over whole range of $\M0$ ($0-3$ TeV). 
We have randomly chosen all the parameters and there 
is no correlation among them. So, these bounds 
depends  only on upper bound of $\msum$. 
From these plots  one can easily  find  the values of individual 
sparticle masses  and the differences 
$\MSL-\MSTAU1$ or  $\MGL-\MST1$ etc.,
which have definite pattern and one can predict the interesting 
possible collider signatures at LHC.

At this large value of $\tan\beta$, $\tilde \tau_1$ can be 
the lightest supersymmetric particle (LSP) in some cases 
depending on the choices  of other parameters, mainly $A_0$
and sign($\mu$). %It should be notable that  
%if we impose $\MSTAU1$ as LSP, then the upper 
%limit on $\tan\beta$ decreases slightly to 62 (53) for
%$\mu > (<) 0$. 
Another consequences of such high values of 
$\TANB$ with positive sign of $\mu$ are
successful explanation of $g-2$ of muon
\cite{Chattopadhyay:2001vx}. 
Again, for positive $\mu$,
$t-b-\tau$ unification is also possible \cite{Baer:2001yy}.

The present global-fit of neutrino data \cite{global2012} after Daya Bay and RENO experiments \cite{dayabayreno} gives $\sin^2\theta_{13} > 0.017$
and $\theta_{13} \approx \theta_C/\sqrt{2}$.               
We have found that a small change in its lower limit leads to a large increase in lower limits
on sparticles masses  ($ \gapp 2$ TeV) from quark-lepton unification, which is the case if recently
discovered boson  at LHC with mass around 125 GeV is the Higgs boson (see fig. \ref{f:smass}).
% if one considers Higgs mass within 
%125 $\pm$ 1 GeV.  
However, since the RGE 
considered for CKM parameters are approximate, we have not
represented the bounds for different lower bounds on $\theta_{13}$.  This would be more precise and reliable when
one considers exact running of CKM parameters. 
%This work is under progress.
%Here, we tried to focus only
%on the correlation between the the upper limit on the sum of neutrino masses and
%the lower limit on sparticle masses in models with EWSB.
%}

As the allowed range of $\tan\beta$ (which is determined mainly to fix $y_\tau$ within an interval) 
is very narrow there appears a definite pattern in the differences between two sparticle masses (see fig. \ref{f:smass}).    
In case of other supersymmetry breaking scenarios one can also expect
similarly strong bounds on sparticle masses from the quark-lepton unification  as one always needs 
large $\tan\beta$ within a narrow range.  
 But, the  differences
in the sparticle masses will then have different definite pattern due to different  
GUT boundary conditions.  The difference between the patterns becomes more prominent 
when the allowed range of $\tan\beta$ is very narrow. This may  make the possibility 
to distinguish different models.
%%%%%%%%%%%%%%%%%%%%%%%%%%%%%%%%%%%%%%%%%%%%%%%%%

%%%%%%%%%%%%%%%%%%%%%%%%%%%%%%%%%%%%%%%%%%%%%%%%%%%%%%%%
%%%%%%%%%%%%%%%%%%%%%%%%%%%%%%%%%%%%%%%%%%%%%%%%%%%%%%%%
%%%%%%%%%%%%%%%%%%%%%%%%%%%%%%%%%%%%%%%%%%%%%%%%%%%%%%%%
%%%%%%%%%%%%%%%%%%%%%%%%%%%%%%%%%%%%%%%%%%%%%%%%%%%%%%%%
%\noindent
\section{Conclusion: }
%%%%%%%%%%%%%%%%%%%%%%%%%%%%%%%%%%%%%%%%%%%%%%%%%
%In conclusion,
%we point out an unique structure of Yukawa matrix (close to unit matrix) at the GUT scale (required
%by quark-lepton unification) among many structures dictated by the different symmetry groups, 
The quark-lepton unification not only
satisfies and/or predicts  all experimental results available till now, but also shows that both results from neutrino experiments and
LHC are complementary.
The quark-lepton unification leads to a strong constraint on the parameter space along with very strong correlations  between the upper limit on $\msum$ and the lower limits on sparticle masses.  This arises due to the fact that there exists a lower limit on $\tan\beta$ for a given  $\msum$ when one demands %Yukawa matrix ($Y_\Delta$) close to unit one to generate 
quasi-degenerate neutrino masses at the GUT scale (which can be generated in GUT models with type II see-saw scenario with an additional family symmetry). As $\msum$ decreases lower limit on $\tan\beta$ increases. For a given high value of $\tan\beta$ there appears very strong lower bounds of sparticle masses
form EWSB. As  $\tan\beta$ increases lower bounds of sparticle masses increase significantly.  We find that $\tan\beta \gapp 55$ is not allowed for $\mu < 0$ (as $\mu^2$ becomes negative and EWSB minima is unstable)   and it constrains $\msum$ $\gapp 1$ eV. 

For $\msum$  $\lapp 1$ eV (constraint from large scale structure of universe) only $\mu > 0$ (which is favored by $(g-2)$ of muon) is allowed and  there exists  strong lower bounds on  sparticle masses $\gapp$ TeV (which are also the bounds from LHC).  A small change in  lower limit of $\theta_{13}$ from zero ($\theta_{13}\approx \theta_C/\sqrt{2}$ after Daya Bay and RENO results)  leads to a large increase in lower limits on sparticles masses  ($ \gapp 2$ TeV) from quark-lepton unification, which is  also the case if recently discovered boson  at LHC with mass around 125 GeV is the Higgs boson. 

%Here, it should be noted that if one considers MSSM instead of CMSSM/mSUGRA, then $m^2_{H_u}$, $m^2_{H_d}$, and other soft mass parameters for squarks and sleptons can take any arbitrary values (satisfed by experimental data) and   
%Recent results from both neutrino experiments and LHC are also complementary in quark-lepton unified models, which dictates the organizing principle for the family structure of Yukawa couplings to produce the observed mass pattern in both lepton and quark sector.

\noindent
{\textbf{Acknowledgements:}}
The author is grateful to Alexei Yu. Smirnov for his valuable comments and suggestions on the manuscript, and  to Xerxes Tata for clarifying the incorporation of threshold corrections to Yukawa couplings in ISASUGRA and their running. The use of the cluster facilities at Harish-Chandra research institute is gratefully acknowledged.

\bibliography{bibtex/nu}

\end{document}